\documentclass[twocolumn,aps,showpacs,preprintnumbers,amsmath,amssymb,unsortedaddress,superscriptaddress]{revtex4}
\usepackage{graphicx,epsf,epsfig,ulem}% Include figure files
\usepackage{bm}% bold math
\usepackage{subfigure}
\usepackage{color}
\usepackage{ulem}
\usepackage{amsmath,amssymb}

\begin{document}

\title{Electrically driven magnetization of diluted magnetic semiconductors actuated by Overhauser effect}
% \shorttitle{Title} %Insert here a short version of the title if it exceeds 70 characters
\bigskip
\author{L. Siddiqui}
\email{lsiddiqu@purdue.edu}
\affiliation{School of Electrical and Computer Engineering, Purdue University, West Lafayette, IN 47907, USA}
\author{A. N. M. Zainuddin}
\affiliation{School of Electrical and Computer Engineering, Purdue University, West Lafayette, IN 47907, USA}
\author{S. Datta}
\affiliation{School of Electrical and Computer Engineering, Purdue University, West Lafayette, IN 47907, USA}
\medskip
\widetext

\begin{abstract}
It is well-known that the Curie temperature, and hence the magnetization, in diluted magnetic semiconductor (DMS) like Ga$_{1-x}$Mn$_x$As can be controlled by changing the equilibrium density of holes in the material. Here, we propose that even with a constant hole density, large changes in the magnetization can be obtained with a relatively small imbalance in the quasi-Fermi levels for up-spin and down-spin electrons. We show, by coupling mean field theory of diluted magnetic semiconductor ferromagnetism with master equations governing the Mn spin-dynamics, that a mere splitting of the up-spin and down-spin quasi-Fermi levels by $0.1$meV will produce the effect of an external magnetic field as large as $1$T as long as the alternative relaxation paths for Mn spins (i.e. spin-lattice relaxation) can be neglected. The physics is similar to the classic Overhauser effect, also called the dynamic nuclear polarization, with the Mn impurities playing the role of the nucleus. We propose that a lateral spin-valve structure in anti-parallel configuration with a DMS as the channel can be used to demonstrate this effect as quasi-Fermi level splitting of such magnitude, inside the channel of similar systems, have already been experimentally demonstrated to produce polarization of paramagnetic impurity spins.
\end{abstract}

\maketitle

\section{Introduction}
\label{intro_sec}
\begin{figure}[]
\begin{center}
\includegraphics[width=0.25\textwidth]{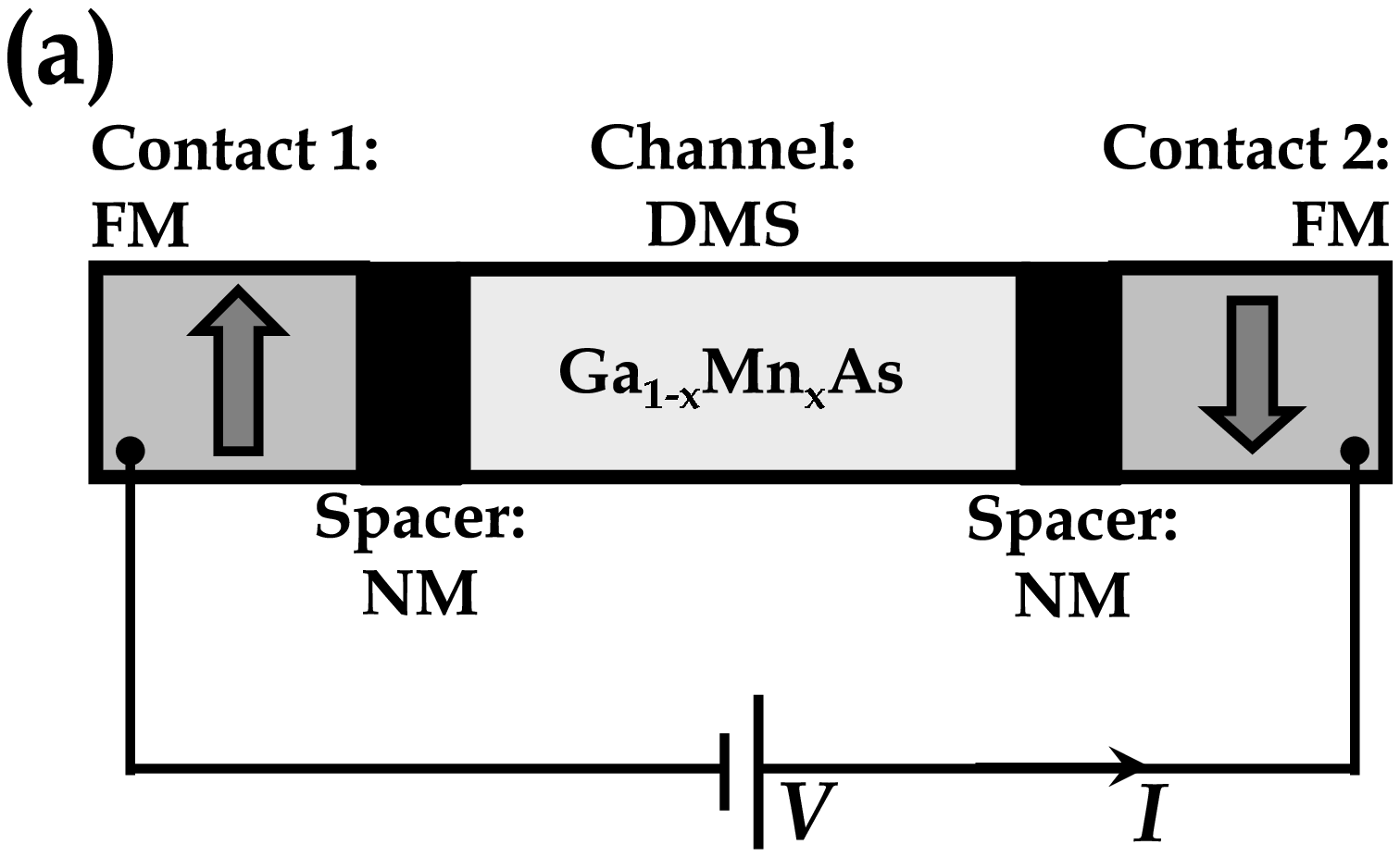}
\end{center}
\begin{center}
\includegraphics[width=0.4\textwidth]{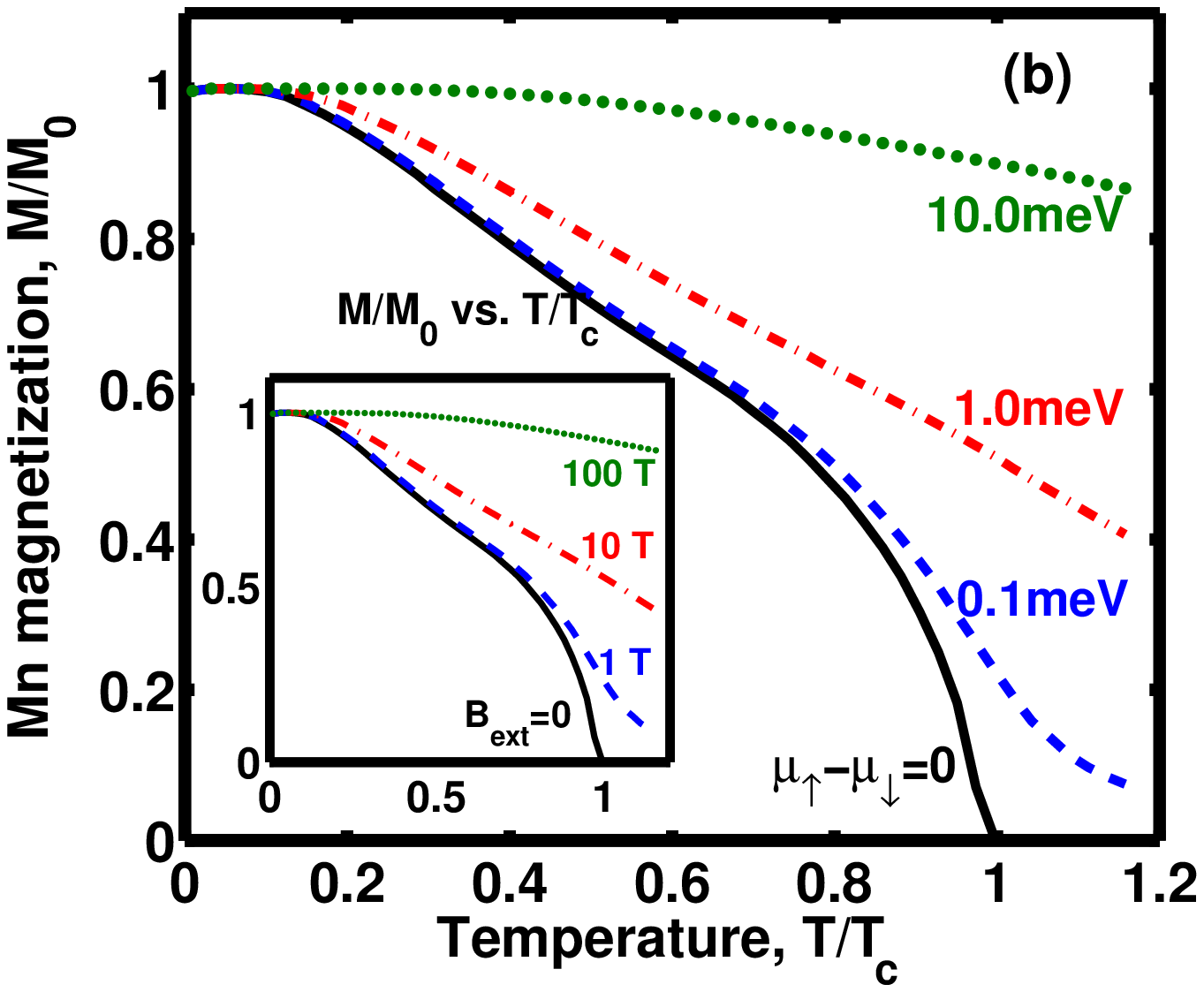}
\end{center}
\begin{center}
\includegraphics[width=0.4\textwidth]{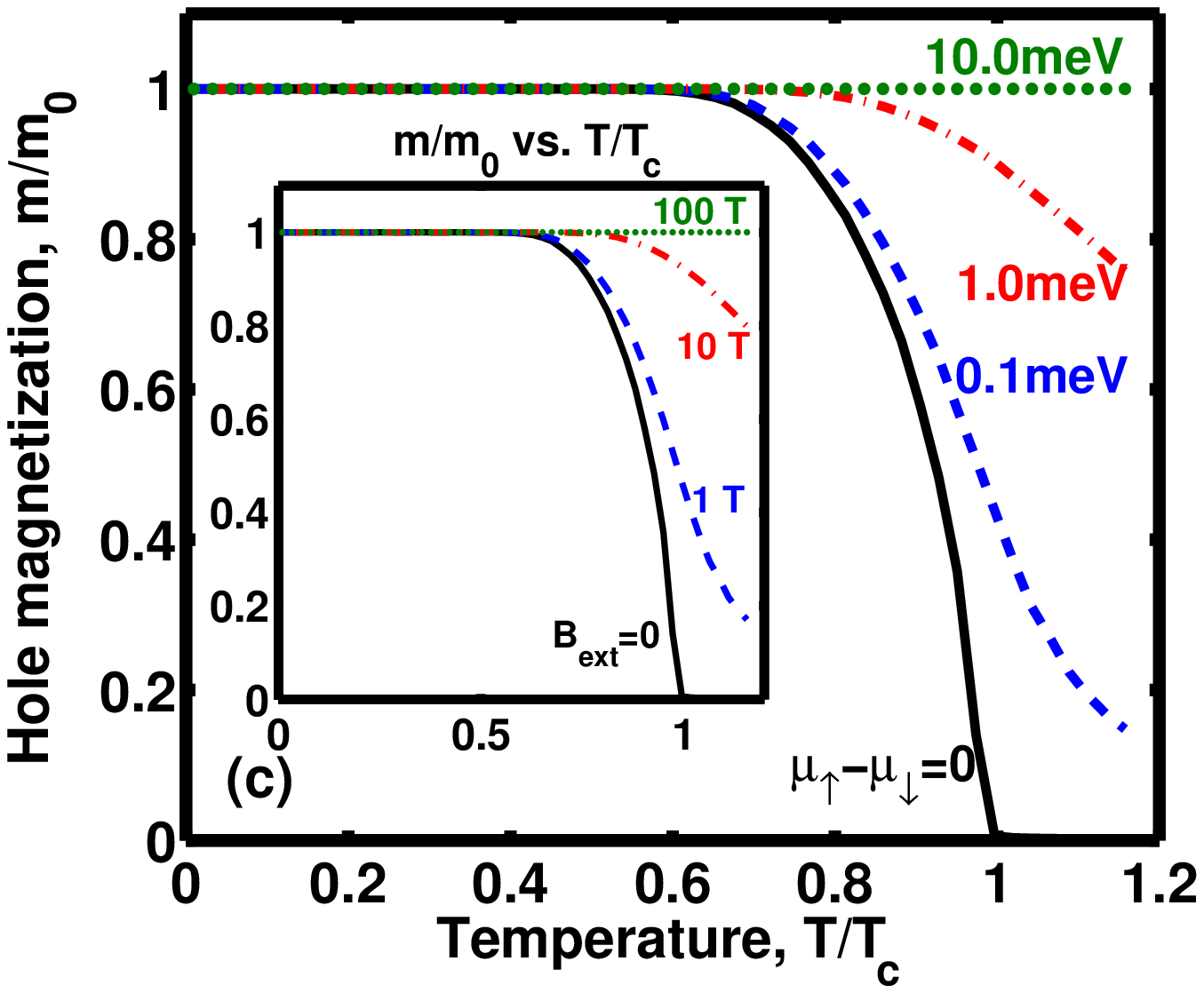}
\end{center}
\caption{(color online). (a) Schematic structure: A dilute magnetic semiconductor (DMS) Ga$_{1-x}$Mn$_{x}$As connected through nonmagnetic (NM) spacers to two ferromagnetic (FM) contact in antiparallel (AP) spin-valve configuration. Under bias $V$, an electronic current $I$, injecting spin-polarized carriers ({\it{holes}}), flows. (b) Mn magnetization $M$, scaled by Mn saturation magnetization $M_0$, vs. temperature $T$, scaled by the Curie temperature $T_c$, of the channel in (a) for different quasi-Fermi level splitting $\mu_{\uparrow}-\mu_{\downarrow}$ ($\mu_{\uparrow(\downarrow)}$ being the channel quasi-Fermi level for up(down)-spin carriers) with external magnetic field $B_{ext}=0$; {\it{inset}}: under different $B_{ext}$ at equilibrium ($\mu_{\uparrow}-\mu_{\downarrow}$=0). (c) Hole magnetization $m$, scaled by saturation hole magnetization $m_0$, vs. temperature $T$, scaled by the Curie temperature $T_c$, of the channel in fig.~\ref{struc_mag}(a): under the same conditions as in fig.~\ref{struc_mag}(b); {\it{inset}}: under the same conditions as in {\it{inset}} of fig.~\ref{struc_mag}(b). The parameter values for these calculations are: $n_{Mn}\sim5.0\times10^{20}$cm$^{-3}$, $n_{h}/n_{Mn}=0.08$, $m^{*}=0.5m_{e}$, $a_0=5.65\AA$, and $J=1$eV.}
\label{struc_mag}
\end{figure}

Electrically driven magnetization of diluted magnetic semiconductors (DMS) has the potentiality to open up new avenues on the map of magneto-electronics and spintronics~\cite{wolf,prinz}. In this regard, electrical manipulation of magnetization has already been demonstrated~\cite{ohno_vcf,boukari_vcf,park_vcf,chiba_vcf_1,nazmul_vcf,chiba_vcf_2,stolichnov_vcf} and theoretically proposed~\cite{nikonov_vcf,ganguly_vcf}. The Curie temperature, in these methods, were controlled by {\it{changing the carrier concentration}} (Fermi level) while keeping {\it{the carrier spin-subsystems in an equilibrium among themselves}}, that is, keeping the quasi-Fermi levels for up-spin ($\mu_{\uparrow}$) and down-spin ($\mu_{\downarrow}$) carriers equal. In contrast, in this work we propose that even with a {\it{constant carrier density}}, large changes in the magnetization can be obtained with a relatively {\it{small imbalance in the spin population}}, that is, a small difference in $\mu_{\uparrow}$ and $\mu_{\downarrow}$. We also propose a structure (fig.~\ref{struc_mag}(a)) for demonstrating the effect that is within current experimental capabilities. In essence, our proposed scheme is similar to the optical manipulation of magnetization in refs.~\cite{oiwa_ocf} and~\cite{nazmul_vcf} where an imbalance in the spin-population is attained by shining circularly polarized light.

Our proposed effect represents a {\it{non-equilibrium}} magnetization resulting from a non-equilibrium bath (the carrier spins) constantly trying to restore equilibrium via spin-flip process due to exchange interaction with localized spins which gets polarized in the process. Indeed the physics is similar to the classic Overhauser effect, also called the dynamic nuclear polarization (DNP)~\cite{dnp}, with the Mn impurities playing the role of the nucleus. To our knowledge this effect has not been employed to actuate non-equilibrium magnetization by electrical excitation although magnetization by optical excitation via, possibly, the same effect~\cite{oiwa_ocf,nazmul_vcf} and demagnetization via the opposite effect~\cite{mag_dyn_expt} has been experimentally observed. Our proposed effect involves electrically driven dynamical polarization of {\it{interacting}} spins (where the polarization of a particular localized spin is affected by the polarization of the neighboring localized spins) and, hence, would be an extension of similar effect studied in the context of {\it{non-interacting}} spins~\cite{feher_overhauser_1,feher_overhauser_2,suhl_overhauser}. Due to this effect a splitting of the up-spin and down-spin quasi-Fermi levels in the channel (fig.~\ref{struc_mag}(a)) by $0.1$meV can have the same effect as an external magnetic field of $1$T (fig.~\ref{struc_mag}(b)). Splitting of this order can be attained by spin-injection into semiconductors~\cite{spin_inj_ref} and has recently been demonstrated in an n-channel GaAs lateral spin-valve operated with anti-parallel contacts to actuate dynamical polarization of non-interacting spin~\cite{dip_prl,lutfe_spinimp}. A similar p-channel Ga$_{1-x}$Mn$_x$As ($x\sim0.05$) structure should be suitable for the demonstration of the proposed effect.

\section{Model Overview}
\label{model_sec}
A number of theoretical papers~\cite{macdonald_mft_suplatt,dsarma_dms_mft,dietl_dms_mft_1,dietl_dms_mft_2,dietl_dms_mft_3,macdonald_mft_1,dsarma_dms_mft_1} have modeled the appearance of ferromagnetic ordering among the Mn ions in Ga$_{1-x}$Mn$_x$As, interacting via the itinerant holes, in terms of a mean field description and explains the experimentally observed~\cite{ohno_apl_dms,potashnik} temperature variation of magnetization in these materials. We adopt exactly the same model as ref.~\cite{dsarma_dms_mft} and have modified it to take into account the non-equilibrium aspect by: {\it{i}}. introducing two different quasi-Fermi levels for up-spin and down-spin holes ($\mu_{\uparrow}$ and $\mu_{\downarrow}$ respctively) inside channel, and {\it{ii}}. writing a master equation to describe the non-equilibrium dynamics of the Mn spins that was used in ref.~\cite{lutfe_spinimp} to semi-quantitatively explain the experimental observation of Mn spin-dynamics in ref.~\cite{dip_prl}. In essence, our model is the same as that in refs.~\cite{mag_dyn_expt,mag_dyn_theo}, which also studies the magnetization dynamics of DMS and uses a more sophisticated valence band description.

With $\mu_{\uparrow}-\mu_{\downarrow}=0$ we get essentially the same results as ref.~\cite{dsarma_dms_mft} (solid curves in figs.~\ref{struc_mag}(b) and~\ref{struc_mag}(c)). Under non-equilibrium situation ($\mu_{\uparrow}-\mu_{\downarrow}\neq 0$), according to our model, we expect to see a strong ferromagnetic ordering among the Mn ions (fig.~\ref{struc_mag}) for moderate values of $\mu_{\uparrow}-\mu_{\downarrow}$ due to reasons that we will discuss later in this paper. A moderate magnitude of $\mu_{\uparrow}-\mu_{\downarrow}\sim 0.1$ meV is quite feasible inside the channel of an anti-parallel lateral spin-valve structure~\cite{dip_prl,lutfe_spinimp} and can be understood in terms of a circuit model presented in ref.~\cite{dip_prl} to explain the experiment therein. We use the same model later in the paper to estimate $\mu_{\uparrow}-\mu_{\downarrow}$.

% \begin{figure}[]
%\begin{center}
% \includegraphics[width=0.35\textwidth]{phole_vs_temp_delmu_Bfield}
% \end{center}
% \caption{\textcolor{red}{MOVED fig.~\ref{carr_pol} to fig.~\ref{struc_mag}(c)} \sout{(color online). Hole magnetization $m$, scaled by saturation hole magnetization $m_0$, vs. temperature $T$, scaled by the Curie temperature $T_c$, of the channel in fig.~\ref{struc_mag}(a): under the same conditions as in fig.~\ref{struc_mag}(b); {\it{inset}}: under the same conditions as in {\it{inset}} of fig.~\ref{struc_mag}(b). The parameter values are also the same as in fig.~\ref{struc_mag}(b).}}
% \label{carr_pol}
% \end{figure}

\section{Theory}
\begin{figure}[]
\begin{center}
\includegraphics[width=0.35\textwidth]{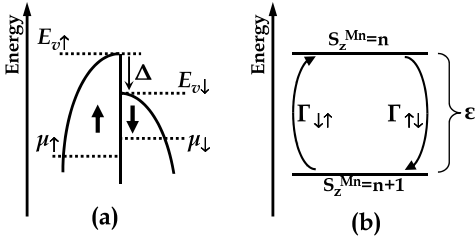}
\end{center}
% \begin{center}
% \includegraphics[width=0.25\textwidth]{mod_self_con}
% \end{center}
\caption{(a) Schematic carrier (hole) bands of the channel in fig.~\ref{struc_mag}(a) for different spins: $E_{v\uparrow(\downarrow)}$, $\mu_{\uparrow(\downarrow)}$ and $\Delta$ being the valence band edges for up(down)-spin holes, their quasi-Fermi levels and their band splitting respectively. (b) Energy levels and relaxation of Mn spins inside the channel in fig.~\ref{struc_mag}(a): $S^{Mn}_{z}$ denotes the six spin states of the $S=5/2$ Mn spins, $\epsilon$ denotes their energy difference, $\Gamma_{\uparrow\downarrow(\downarrow\uparrow)}$ denotes the rate of transition from lower (higher) z-component state to a higher (lower) z-component state and, $n=-5/2$, $-3/2$, $\hdots$, $3/2$.}
\label{band_level_model}
\end{figure}

% \section{Nonequilibrium magnetization}
\label{theory_sec}
The spontaneous ferromagnetic ordering of the localized Mn spins in a DMS material arises due to the hole mediated exchange interaction between them. In the context of mean field theory of DMS~\cite{dsarma_dms_mft}, the carriers `feel' an exchange field due to the polarized Mn spins in addition to any external magnetic field $B_{ext}$, which separate the up-spin band from the down-spin band (fig.~\ref{band_level_model}(a)) in energy by,
\begin{eqnarray}
&& \Delta=\Delta_{(ex)}+g_{h}\mu_BB_{ext} \label{delta} \\
&& \Delta_{(ex)}=Ja^3_0n_{Mn}\langle S^{Mn}_z\rangle \nonumber
\end{eqnarray} where, $a_0$ is the lattice constant, $g_{h}$ is the g-factor of the carrier (hole), $\mu_B$ is the Bohr magneton, $\Delta_{(ex)}$ is the separation between up-spin band and down-spin band due to exchange field and $\langle S^{Mn}_{z}\rangle$ is the average z-component of $S=5/2$ Mn spins:
\begin{eqnarray}
% P_{Mn}&=& \frac{\langle S_z^{Mn} \rangle}{5/2} \label{PMndef}\\
\langle S_z^{Mn} \rangle&=&\sum_ssF_s \label{SzMndef}
%\frac{5}{2}F_{+5/2}+\frac{3}{2}F_{+3/2}+\frac{1}{2}F_{+1/2} \nonumber\\
% &-&\frac{1}{2}F_{1/2}-\frac{3}{2}F_{3/2}-\frac{5}{2}F_{5/2} \label{SzMndef}
\end{eqnarray} $F_s$ being the probability of a Mn spin being in $S^{Mn}_z=s$ state ($s=5/2$, $3/2$, $\hdots$, $-5/2$).

The carrier band splitting in addition to splitting of quasi-Fermi levels for different carrier spins (fig.~\ref{band_level_model}(a)) lead to a non-zero average z-component of hole-spin $\langle S^h_z\rangle$ due to unequal carrier concentrations for up-spin ($n_{h,\uparrow}$) and down-spin ($n_{h,\downarrow}$), which are given by:
\begin{eqnarray}
% P_h&=&\frac{n_{h,\uparrow}-n_{h,\downarrow}}{n_{h,\uparrow}+n_{h,\downarrow}}\label{phole}\\
\langle S^h_z\rangle &=&\frac{1}{2}\frac{n_{h,\uparrow}-n_{h,\downarrow}}{n_{h,\uparrow}+n_{h,\downarrow}}\label{phole}\\
n_{h,\uparrow(\downarrow)}&=& \int dE(1-f_{\uparrow(\downarrow)}(E))D_{\uparrow(\downarrow)}(E)\label{nholeupdn}
\end{eqnarray} where, $f_{\uparrow(\downarrow)}(E)$ is Fermi function with Fermi level $\mu_{\uparrow(\downarrow)}$ and temperature $T$: $f_{\uparrow(\downarrow)}(E)=\left[1+\exp\left\{\left(E-\mu_{\uparrow(\downarrow)}\right)/k_BT\right\}\right]^{-1}$, and $D_{\uparrow(\downarrow)}$ is the three dimensional density of states for up(down)-spin carriers calculated assuming a parabolic band with an effective mass of $m^*$ and a band-edge at $E_{v\uparrow(\downarrow)}$ (fig.~\ref{band_level_model}(a)). $E_{v\uparrow}$, $E_{v\downarrow}$ and $\Delta$ are related by $E_{v\downarrow}-E_{v\uparrow}=\Delta$ and the charge neutrality condition: $n_{h}=n_{h,\uparrow}+n_{h,\downarrow}$.

The Mn spins on the other hand feel an exchange field due to the spin-polarized carriers in addition to any external magnetic field, which lead to energy level splitting (fig.~\ref{band_level_model}(b)),
\begin{eqnarray}
&& \epsilon=\epsilon_{(ex)}+g_{Mn}\mu_BB_{ext} \label{Mnsplit} \\
&& \epsilon_{(ex)}=Ja^3_0n_{h}\langle S^h_{z}\rangle \nonumber
\end{eqnarray} (a new variable $\epsilon_{(ex)}$ appears in the above equation) where, $g_{Mn}$ is the g-factor of Mn spins and $\epsilon_{(ex)}$ is the splitting of Mn spin levels due to exchange field.

Up to this point, all the equations and quantities, except different quasi-Fermi levels $\mu_{\uparrow}$ and $\mu_{\downarrow}$, are essentially the same as ref.~\cite{dsarma_dms_mft} and takes into account the non-equilibrium effects of Mn spin-polarization on the holes. To consider the non-equilibrium effects of the holes on the Mn spins and the resulting non-equilibrium dynamics we solve the dynamic rate equation, which was also used in ref.~\cite{lutfe_spinimp,mag_dyn_theo}, for the occupation probabilities $F_s$:
\begin{eqnarray}
\frac{d}{dt}\left[\begin{array}{c} F_{-5/2}\\
F_{-3/2}\\
F_{-1/2}\\
F_{+1/2}\\
F_{+3/2}\\
F_{+5/2} \end{array}\right]={\mathbf{\Gamma}}\left[\begin{array}{c}F_{-5/2}\\
F_{-3/2}\\
F_{-1/2}\\
F_{+1/2}\\
F_{+3/2}\\
F_{+5/2}
\end{array}\right] \label{RateEq}
\end{eqnarray} at steady state (by setting $dF_s/dt=0$) under the normalization constraint: $\sum_{s}F_s=1$ where,
\begin{widetext}
\begin{eqnarray}
\label{Req}
{\mathbf{\Gamma}}=\left[ \begin{array}{cccccc}
-\Gamma_{\uparrow\downarrow} & \Gamma_{\downarrow\uparrow} & 0 & 0 & 0 & 0\\
\Gamma_{\uparrow\downarrow} & -(\Gamma_{\uparrow\downarrow}+\Gamma_{\downarrow\uparrow}) & \Gamma_{\downarrow\uparrow} & 0 & 0 & 0\\
0 & \Gamma_{\uparrow\downarrow} & -(\Gamma_{\uparrow\downarrow}+\Gamma_{\downarrow\uparrow}) & \Gamma_{\downarrow\uparrow} & 0 & 0\\
0 & 0 & \Gamma_{\uparrow\downarrow} & -(\Gamma_{\uparrow\downarrow}+\Gamma_{\downarrow\uparrow}) & \Gamma_{\downarrow\uparrow} & 0\\
0 & 0 & 0 & \Gamma_{\uparrow\downarrow} & -(\Gamma_{\uparrow\downarrow}+\Gamma_{\downarrow\uparrow}) & \Gamma_{\downarrow\uparrow}\\
0 & 0 & 0 & 0 & \Gamma_{\uparrow\downarrow} & -\Gamma_{\downarrow\uparrow}\\
\end{array} \right]
\end{eqnarray}
\end{widetext} $\Gamma_{\uparrow\downarrow}$ and $\Gamma_{\downarrow\uparrow}$ (fig.~\ref{band_level_model}(b)), as valence band holes surrounding the Mn spins act as their spin-bath and also as Mn spins have negligible spin-lattice relaxation rate in comparison~\cite{mag_dyn_expt}, are given by:
\begin{subequations}
\begin{eqnarray}
\Gamma_{\uparrow\downarrow}=\frac{2\pi}{h}J^2\int dE D_\downarrow(E+\epsilon)(1-f_\downarrow(E+\epsilon))D_\uparrow(E)f_\uparrow(E)&& \nonumber \\
&& \label{gam_ud}\\
\Gamma_{\downarrow\uparrow}=\frac{2\pi}{h}J^2\int dE D_\uparrow(E)(1-f_\uparrow(E))D_\downarrow(E+\epsilon)f_\downarrow(E+\epsilon)&& \nonumber \\
&& \label{gam_du}
\end{eqnarray}
\label{tranrates}
\end{subequations}

\section{Results: nonequilibrium magnetization}
\label{result_sec}
\begin{figure}[]
\begin{center}
\includegraphics[width=0.3\textwidth]{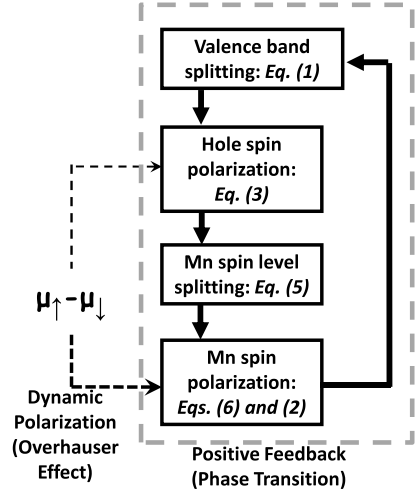}
\end{center}
% \begin{center}
% \includegraphics[width=0.25\textwidth]{mod_self_con}
% \end{center}
\caption{Nonequilibrium magnetization arises from an interplay of dynamical spin-polarization (Overhauser effect) that is driven by the quasi-Fermi level splitting $\mu_{\uparrow}-\mu_{\downarrow}$ between up-spin and down-spin holes and a positive feedback that is responsible for equilibrium magnetization below Curie temperature}
\label{overhauser_pos_fb}
\end{figure}

The eqs.~\ref{phole},~\ref{Mnsplit},~\ref{RateEq},~\ref{SzMndef} and~\ref{delta}, are solved sequentially and self-consistently (fig.~\ref{overhauser_pos_fb}) to calculate the Mn spin polarization and hole polarization in figs.~\ref{struc_mag}(b) and~\ref{struc_mag}(c) respectively. Magnetization of the Mn and holes inside channel material (Ga$_{1-x}$Mn$_x$As) is then calculated from: $M=2M_0\langle S^{Mn}_z\rangle/5$ and $m=2m_0\langle S^h_z\rangle$ respectively, where the corresponding saturation magnetizations, $M_0=5g_{Mn}\mu_Bn_{Mn}/2$ and $m_o=g_h\mu_Bn_h/2$. The iterative loop in fig.~\ref{overhauser_pos_fb} embodies a positive feedback loop that gives rise to equilibrium magnetization below Curie temperature in such materials. Such positive feedback in combination with Overhauser effect (fig.~\ref{overhauser_pos_fb}) is what gives rise to non-equilibrium magnetization even above Curie temperature. The results of the calculations, shown in figs.~\ref{struc_mag}(b) and~\ref{struc_mag}(c), use realistic parameter values~\cite{dsarma_dms_mft} and gives equilibrium magnetization characteristics (calculated by setting $\mu_{\uparrow}-\mu_{\downarrow}=0$) similar to experimental observations~\cite{ohno_apl_dms,potashnik}. Two sets of calculations were performed: 1) calculations for different values of $\mu_{\uparrow}-\mu_{\downarrow}$ by setting $B_{ext}=0$, whose results are shown in the main plots of figs.~\ref{struc_mag}(b) and~\ref{struc_mag}(c), and 2) calculations for different values of $B_{ext}$ by setting $\mu_{\uparrow}-\mu_{\downarrow}=0$ (equilibrium), whose results are shown in the inset plots of figs.~\ref{struc_mag}(b) and~\ref{struc_mag}(c). By comparing the main plots in figs.~\ref{struc_mag}(b) and~\ref{struc_mag}(c) with their corresponding inset plots, we observe that the values of $\mu_{\uparrow}-\mu_{\downarrow}$ corresponding to the curves in the main plots maintain a proportionality relation with the values of $B_{ext}$ corresponding to the similar curves in the corresponding inset plots ($0.1$meV$:1.0$meV$:10.0$meV$=1$T$:10$T$:100$T). Such proportionality relation is, by no means, a coincidence and is maintained in the calculations done with different values of $J$, $n_{Mn}$, and $n_h$ (results not shown in this paper). Although the Curie temperature, the saturation magnetization and the shapes of the magnetization vs. temperature curves are different for different values of $J$, $n_{Mn}$, and $n_h$ due to the dependence of the exchange field on these parameters (eqs.~\ref{delta} and~\ref{Mnsplit}) the strength of the effect remains the same, i.e. the magnetizations for a quasi-Fermi level splitting of $~0.1$meV without any external magnetic field, when alternative Mn spin relaxation paths can be neglected, is equal to the magnetizations for an external magnetic field of $1$T at equilibrium. As a result, changing $J$, $n_{Mn}$, and $n_h$ by changing the doping, changing the Mn mole fraction, introducing disorder~\cite{lipinska} or by using a different DMS material (that has itinerant carrier mediated exchange interaction of localized spins) will not play any significant role as far as the strength of the effect is concerned for reasons to be discussed in the next section.

% \begin{equation}
% M = g_{Mn}\mu_Bn_{Mn}\langle S_z^{Mn}\rangle
% \label{magnetization}
% \end{equation} where $n_{Mn}$ is the concentration of $Mn$ atoms. The reults of the calculations, shown in figs.~\ref{struc_mag}(b) and~\ref{carr_pol}, uses realistic parameter values~\cite{dsarma_dms_mft} and gives equilibrium magnetization characteristics (calculated by setting $\mu_{\uparrow}-\mu_{\downarrow}=0$) similar to experimental observations~\cite{ohno_apl_dms,potashnik}.

\section{Discussion}
\label{discuss_sec}
The results in figs.~\ref{struc_mag}(b) and~\ref{struc_mag}(c) can be anticipated if the effect of $\mu_{\uparrow}-\mu_{\downarrow}$ is considered as an effective external magnetic field as far as the hole spin-polarization and Mn spin-polarization are concerned. We can show that the functional dependence of $\langle S^h_z\rangle$ on $\Delta$ and $\mu_{\uparrow}-\mu_{\downarrow}$ obeys the following relation (see appendix~\ref{hole_spin_app}):
\begin{eqnarray}
\langle S^h_z\rangle(\Delta,\mu_{\uparrow}-\mu_{\downarrow}) &=& \langle S^h_z\rangle(\Delta+\mu_{\uparrow}-\mu_{\downarrow},0)
\label{hole_spin_eq}
\end{eqnarray} The first term is the hole spin-polarization for a valence band splitting of $\Delta$ and a non-zero quasi-Fermi level splitting $\mu_{\uparrow}-\mu_{\downarrow}$ while the second term corresponds to the hole spin-polarization for an additional valence band splitting of $\mu_{\uparrow}-\mu_{\downarrow}$ over the orginal value $\Delta$ and a zero quasi-Fermi level splitting (equilibrium). At this point, one can observe that the results in the main plot (i.e. for $B_{ext}=0$) and the inset plot (i.e. for $\mu_{\uparrow}-\mu_{\downarrow}=0$) of fig.~\ref{struc_mag}(c) (i.e. hole spin-polarization) correspond to $\langle S^h_z\rangle(\Delta_{(ex)},\mu_{\uparrow}-\mu_{\downarrow})$ i.e. $\langle S^h_z\rangle(\Delta_{(ex)}+\mu_{\uparrow}-\mu_{\downarrow},0)$ (eq.~\ref{hole_spin_eq}) and $\langle S^h_z\rangle(\Delta_{(ex)}+g_h\mu_BB_{ext},0)$ respectively. The mathematical equivalence of the last two expressions suggests that, as far as hole spin-polarization is concerned, $\mu_{\uparrow}-\mu_{\downarrow}$ has the equivalent effect of an effective external magnetic field of $B^{h}_{ext}=(\mu_{\uparrow}-\mu_{\downarrow})/(g_{h}\mu_{B})$. On the other hand, for Mn spins it can also be shown that the functional dependence of $\langle S^{Mn}_z\rangle$ on $\epsilon$ and $\mu_{\uparrow}-\mu_{\downarrow}$ obeys the following relation (see appendix~\ref{Mn_spin_app}):
\begin{eqnarray}
\langle S^{Mn}_z\rangle(\epsilon,\mu_{\uparrow}-\mu_{\downarrow}) &=& \langle S^{Mn}_z\rangle(\epsilon+\mu_{\uparrow}-\mu_{\downarrow},0)
\label{Mn_spin_eq}
\end{eqnarray} In this case, the first term is the Mn spin-polarization for an energy level splitting of $\epsilon$ and a non-zero quasi-Fermi level splitting of $\mu_{\uparrow}-\mu_{\downarrow}$ while the second term corresponds to the Mn spin-polarization for an additional energy level splitting of $\mu_{\uparrow}-\mu_{\downarrow}$ over the orginal value $\epsilon$ and a zero quasi-Fermi level splitting (equilibrium). At this point, one can observe that the results in the main plot (i.e. for $B_{ext}=0$) and the inset plot (i.e. for $\mu_{\uparrow}-\mu_{\downarrow}=0$) of fig.~\ref{struc_mag}(b) (i.e. Mn spin-polarization) correspond to $\langle S^{Mn}_z\rangle(\epsilon_{(ex)},\mu_{\uparrow}-\mu_{\downarrow})$ i.e. $\langle S^{Mn}_z\rangle(\epsilon_{(ex)}+\mu_{\uparrow}-\mu_{\downarrow},0)$ (eq.~\ref{Mn_spin_eq}) and $\langle S^{Mn}_z\rangle(\epsilon_{(ex)}+g_h\mu_BB_{ext},0)$ respectively. The mathematical equivalence of the last two expressions shows that, as far as Mn spin-polarization is concerned, $\mu_{\uparrow}-\mu_{\downarrow}$ has the equivalent effect of an effective external magnetic field of $B^{Mn}_{ext}=(\mu_{\uparrow}-\mu_{\downarrow})/(g_{Mn}\mu_{B})$ acting on the Mn spins at equilibrium. For $g_h\sim g_{Mn}\sim 2$ the equivalent external magnetic fields for hole spins and Mn spins ($B^{h}_{ext}$ and $B^{Mn}_{ext}$), that are mentioned above, would be eqaul and be given by
\begin{equation}
B^{eff}_{ext}\approx\frac{\mu_{\uparrow}-\mu_{\downarrow}}{2\mu_B}\label{equiBfield}
\end{equation} Herein lies the strength of the effect: a mere difference of $0.1$meV between $\mu_{\uparrow}$ and $\mu_{\downarrow}$ is strong enough to produce the effect corresponding to that of an external magnetic field as large as $1$T. One can notice that the arguments presented above do not depend on $J$, $n_{Mn}$ or $n_{h}$ and require that the alternative relaxation paths for the Mn spins can be neglected (which entered through the neglect of spin-lattice relaxation rate while writing down the eqs.~\ref{tranrates} and was subsequently used in the derivation of eq.~\ref{Mn_spin_eq}). {\it{As a result, the strength of the effect (eq.~\ref{equiBfield}) is insensitive to a change in $J$, $n_{Mn}$, and $n_h$, and, hence, is relatively insensitive to our choice of mean field (eqs.~\ref{delta} and~\ref{Mnsplit}) as long as the alternative relaxation paths for the Mn spins can be neglected}}. Since 1) our central result (eq.~\ref{equiBfield}), originating from non-equilibrium magnetization dynamics, is relatively insensitive to our choice of mean field (eqs.~\ref{delta} and~\ref{Mnsplit}), 2) eqs.~\ref{delta} and~\ref{Mnsplit} describe the equilibrium temperature dependence of magnetization, at least, qualitatively~\cite{dsarma_dms_mft}, and 3) a model~\cite{mag_dyn_theo}, which essentially uses our choice of mean field, has been used to successfully explain recent experiments~\cite{mag_dyn_expt} on non-equilibrium magnetization dynamics we feel justified in leaving it to future work to assess the need for improving eqs.~\ref{delta} and~\ref{Mnsplit}.

\begin{figure}[]
\begin{center}
\includegraphics[width=0.35\textwidth]{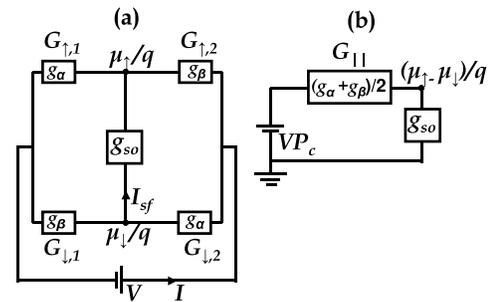}
\end{center}
\caption{(a) Circuit model of the strcuture in fig.~\ref{struc_mag}(a) (adopted from ref.~\cite{dip_prl}): $g_{so}$, $G_{\uparrow(\downarrow),1}$, $G_{\uparrow(\downarrow),2}$, $g_{\alpha}$ and, $g_{\beta}$ are the carrier spin-flip conductance independent of Mn spins, contact 1 conductance for up(down)-spin, contact 2 conductance for up(down)-spin, majority spin conductance and, minority spin conductance respectively. (b) Simplified circuit diagram of the model circuit in (a): $G_{\|}$ and $P_{c}=(g_\alpha-g_\beta)/(g_\alpha+g_\beta)$ are the terminal conductance for parallel (P) configuration and the contact polarization respectively.}
\label{eqckt_delmu}
\end{figure}

Experimental realization of quasi-Fermi level splitting, similar to the values mentioned above, has already been demonstrated to drive dynamical polarization of non-interacting spins~\cite{dip_prl,lutfe_spinimp} with a structure similar to the one shown in fig.~\ref{struc_mag}(a). One can quite legitimately envision that with further improvement of spin-injection process in terms of contact polarization $P_{c}$ and contact conductance and hence the parallel terminal conductance $G_{\|}$ one can achieve even higher $\mu_{\uparrow}-\mu_{\downarrow}$ leading to higher degree of ferromagnetic ordering that would otherwise require immensely large magnetic field (fig.~\ref{struc_mag}). The material property of the DMS that acts against attaining quasi-fermi level splitting between up-spin and down-spin holes is the spin lifetime of valence band holes $\tau_{so}$ independent of Mn that give rise to the spin-flip conductance $g_{so}$. The effect of all these ingredients of a spin-valve structure on $\mu_{\uparrow}-\mu_{\downarrow}$ can be concisely pictured in terms of the circuit model in fig.~\ref{eqckt_delmu}(a) which we have adopted from ref.~\cite{dip_prl} and is valid for a channel length in fig.~\ref{struc_mag}(a) that is smaller than the spin-diffusion length. Since the spin-diffusion length of the magnetic semiconductors has not been reported in the literature to the best of our knowledge we are unable to conclusively comment, as far as the channel length is concerned, on the scope of the analysis to follow and will limit our analysis to the thinnest possible (2D) channel having a thickness of atomic dimensions in the transport direction. Nevertheless, it will touch upon some key ingredients that affect $\mu_{\uparrow}-\mu_{\downarrow}$. Moreover, the derivation of eqs.~\ref{hole_spin_eq} and~\ref{Mn_spin_eq} do not rely on the shape of the density of states as long as the density of states for both up-spin and down-spin carriers have the same energy dependence (so that one can write: $D_{\uparrow}(E)=D_{\downarrow}(E+\Delta)$). As a result, as far as the strength of the effect is concerned, whether the channel is 2D or 3D does not play any significant role.

Upon simplification of that circuit model as shown in fig.~\ref{eqckt_delmu}(b) we get: $\mu_{\uparrow}-\mu_{\downarrow}=qVP_{c}(1+\frac{g_{so}}{G_{\|}})^{-1}$ where, $V$ is the applied bias and $q$ is the electronic charge. We estimate $g_{so}\sim10^{10}\Omega^{-1}$m$^{-2}$ for Ga$_{1-x}$Mn$_x$As from the relation $g_{so}=\frac{q^2}{h}D\frac{\hbar}{\tau_{so}}$ using $\tau_{so}=1$ps (spin life-time of GaAs valence band holes~\cite{kainz}) and a 2D density of states value $D\sim10^{37}$J$^{-1}$m$^{-2}$ (estimated using the valence band effective mass used in fig.~\ref{struc_mag}(b), which is of the same order of magnitude as the number calculated from 3D density of states for a thickness of atomic dimensions ($\sim 1$nm)). For such value of $g_{so}$ if we use the terminal conductance and contact polarization values of refs.~\cite{dip_prl,lutfe_spinimp} ($G_{\|}\sim10^7\Omega^{-1}$m$^{-2}$, $P_c\sim0.5$) we estimate $\mu_{\uparrow}-\mu_{\downarrow}\sim1$meV for an applied voltage of $V=1$V. However, for a $\tau_{so}=10$fs (the value used in ref.~\cite{mag_dyn_expt}) one would have to increase $G_{\|}$ to $\sim10^9\Omega^{-1}$m$^{-2}$ (similar to the values in ref.~\cite{koo_sv_07}) to get the same effect.

% \begin{eqnarray}
% \frac{\mu_{\uparrow}-\mu_{\downarrow}}{qVP_{c}}&=& \frac{1}{1+\frac{g_{so}}{G_{\|}}}\label{musplit}
% \end{eqnarray}

\section{Summary}
\label{summary_sec}
In summary, we have proposed a novel mechanism of attaining and controlling ferromagnetic ordering in DMS materials. We argue that the non-equilibrium accumulation of carrier spins drives the spin alignment of magnetic impurities in DMS via the exchange interaction which is responsible for the origin of the ferromagnetism in such materials in the first place. We find the effect to be quite strong when we consider that such degree of ferromagnetic ordering would otherwise have to be attained by applying a very large magnetic field and, also that this strength is insensitive to a change in several crucial parameters (magnetically active Mn concentration, hole concentration, and strength of exchange interaction between Mn spins and valence band holes) that describe the equilibrium ferromagnetism of such systems. We believe that, with the existing experimental sophistication achieved over the years, our proposed scheme would be realized in the near future and will usher in a new paradigm in the experimental investigation of ferromagnetic phase transition and also in the applications benefitting from the strong control of magnetism such as the magnetocaloric applications~\cite{tishin}.

This work is supported by the Office of Naval Research under Grant No. N00014-06-1-0025.

\appendix
\section{Derivation of eq.~\ref{hole_spin_eq}}
\label{hole_spin_app}
\noindent Starting from eqs.~\ref{phole} and~\ref{nholeupdn} we get
\begin{eqnarray}
&& \langle S^h_z\rangle(\Delta,\mu_{\uparrow}-\mu_{\downarrow})=\frac{1}{2n_h}\left[\int dE\left\{D_{\uparrow}(E)-D_{\downarrow}(E)\right\}\right. \nonumber \\
&-& \left.\int dE\left\{f_{\uparrow}(E)D_{\uparrow}(E)-f_{\downarrow}(E)D_{\downarrow}(E)\right\}\right]
\label{eq101}
\end{eqnarray} where $n_h$ was assumed to remain constant (varying with neither $\Delta$ nor $\mu_{\uparrow}-\mu_{\downarrow}$) on the ground of charge neutrality. Now, for a given valence band splitting $\Delta$ and a given quasi-Fermi level splitting $\delta\mu\equiv\mu_{\uparrow}-\mu_{\downarrow}$ (fig.~\ref{band_level_model}) we can write
\begin{eqnarray}
D_{\uparrow}(E) &=& D_{\downarrow}(E+\Delta)\equiv D(E) \label{eq201} \\
f_{\uparrow}(E) &=& f_{\downarrow}(E-\delta\mu)\equiv f(E) \label{eq301}
\end{eqnarray} Substituting the above results into eq.~\ref{eq101}
\begin{eqnarray}
&& \langle S^h_z\rangle(\Delta,\mu_{\uparrow}-\mu_{\downarrow})=\frac{1}{2n_h}\left[\int dE\left\{D(E)-D(E-\Delta)\right\}\right. \nonumber \\
&-& \left. \int dE\left\{f(E)D(E)-f(E+\delta\mu)D(E-\Delta)\right\}\right]
\label{eq401}
\end{eqnarray} The part
\begin{eqnarray}
&& \int dE\left\{D(E)-D(E-\Delta)\right\} \nonumber \\
&=& \int dE D(E)-\int dE D(E-\Delta) \nonumber \\
&=& \int dE D(E)-\int dE' D(E') \nonumber
\end{eqnarray} by performing a change of variable on the second integration in the previous step. It, finally, leads to
\begin{equation}
\int dE\left\{D(E)-D(E-\Delta)\right\}=0 \nonumber
\end{equation} Substituting the above result in eq.~\ref{eq401} we get
\begin{eqnarray}
\langle S^h_z\rangle(\Delta,\mu_{\uparrow}-\mu_{\downarrow}) &=& \frac{1}{2n_h}\int dE\left\{f(E)D(E)\right. \nonumber \\
&-& \left. f(E+\delta\mu)D(E-\Delta)\right\}
\label{eq501}
\end{eqnarray} Proceeding further,
\begin{eqnarray}
&& \langle S^h_z\rangle(\Delta,\mu_{\uparrow}-\mu_{\downarrow}) \nonumber \\
&=& \frac{1}{2n_h}\left\{\int dE f(E)D(E)-\int dE f(E+\delta\mu)D(E-\Delta)\right\} \nonumber \\
% &=& \frac{1}{2n_h}\left\{\int dE f(E)D(E)-\int dE f(E+\delta\mu)D(E-\Delta)\right\} \nonumber\\
&=& \frac{1}{2n_h}\left\{\int dE f(E)D(E)-\int dE' f(E')D(E'-\Delta-\delta\mu)\right\} \nonumber
\end{eqnarray} by performing a change of variable on the second integration in the previous step. Finally, it leads to
\begin{eqnarray}
\langle S^h_z\rangle(\Delta,\mu_{\uparrow}-\mu_{\downarrow}) &=& \frac{1}{2n_h}\int\left\{dE f(E)D(E)\right. \nonumber \\
&-& \left. f(E)D(E-\Delta-\delta\mu)\right\}
\label{eq601}
\end{eqnarray} while substituting $\mu_{\uparrow}-\mu_{\downarrow}\equiv\delta\mu=0$ and $\Delta=\Delta'+\delta\mu$ in eq.~\ref{eq501} we find
\begin{eqnarray}
\langle S^h_z\rangle(\Delta'+\delta\mu,0) &=& \frac{1}{2n_h}\int dE\left\{f(E)D(E)\right. \nonumber \\
&-& \left. f(E)D(E-\Delta'-\delta\mu)\right\}
% \label{eq701}
\end{eqnarray} which, trivially, leads to
\begin{eqnarray}
\langle S^h_z\rangle(\Delta+\mu_{\uparrow}-\mu_{\downarrow},0) &=& \frac{1}{2n_h}\int dE\left\{f(E)D(E)\right. \nonumber \\
&-& \left. f(E)D(E-\Delta-\delta\mu)\right\}
\label{eq701}
\end{eqnarray} The right hand sides of eqs.~\ref{eq601} and~\ref{eq701} being equal, we equate their left hand sides and arrive at the results in eq.~\ref{hole_spin_eq}
% \begin{equation}
% \langle S^h_z\rangle(\Delta,\mu_{\uparrow}-\mu_{\downarrow})=\langle S^h_z\rangle(\Delta+\mu_{\uparrow}-\mu_{\downarrow},0)
% \end{equation}

\section{Derivation of eq.~\ref{Mn_spin_eq}}
\label{Mn_spin_app}
Starting from eqs.~\ref{RateEq} and~\ref{Req}, at steady state, we get
\begin{eqnarray}
\frac{F_{-5/2}}{F_{-3/2}}=\frac{F_{-3/2}}{F_{-1/2}}=\cdots=\frac{F_{+3/2}}{F_{+5/2}}=\alpha
\label{eq150}
\end{eqnarray} where, we have defined
\begin{eqnarray}
\alpha\equiv\frac{\Gamma_{\downarrow\uparrow}}{\Gamma_{\uparrow\downarrow}}
\label{eq250}
\end{eqnarray} From eqs.~\ref{eq150} and probability conservation: $\sum_{s}F_s=1$ we get
\begin{eqnarray}
\frac{F_{-5/2}}{\alpha^5}=\frac{F_{-3/2}}{\alpha^4}=\cdots=F_{+5/2}=\frac{1-\alpha}{1-\alpha^{6}}
\label{eq350}
\end{eqnarray} Substituting the above results in eq.~\ref{SzMndef} we find
\begin{eqnarray}
\langle S^{Mn}_z\rangle=\frac{\left(1-\alpha\right)^2}{2\left(1-\alpha^6\right)}\left(5\alpha^4+8\alpha^3+9\alpha^2+8\alpha+5\right)
\label{eq450}
\end{eqnarray} which shows that $\langle S^{Mn}_z\rangle$ {\it{at steady state is solely a function of}} $\alpha$. Now, from eqs.~\ref{tranrates}, \ref{eq201}, \ref{eq301} and~\ref{eq250}
\begin{eqnarray}
\alpha=\frac{\int dE D(E)\left\{1-f(E)\right\}D(E+\epsilon-\Delta)f(E+\epsilon+\delta\mu)}{\int dE D(E+\epsilon-\Delta)\left\{1-f(E+\epsilon+\delta\mu)\right\}D(E)f(E)} && \nonumber \\
&& 
\label{eq550}
\end{eqnarray} For a given value of $E$ the ratio of the integrands in the above equation
\begin{eqnarray}
&& \frac{D(E)\left\{1-f(E)\right\}D(E+\epsilon-\Delta)f(E+\epsilon+\delta\mu)}{D(E+\epsilon-\Delta)\left\{1-f(E+\epsilon+\delta\mu)\right\}D(E)f(E)} \nonumber \\
&=& \left\{\frac{1-f(E)}{f(E)}\right\}\left\{\frac{1-f(E+\epsilon+\delta\mu)}{f(E+\epsilon+\delta\mu)}\right\}^{-1} \nonumber \\
&=& \exp\left\{-\frac{\epsilon+\delta\mu}{k_BT}\right\} \nonumber
\end{eqnarray} Using the above result while treating the integrations in eq.~\ref{eq550} as summation over energy and making use of the identity
\begin{eqnarray}
\frac{N_1}{D_1}=\frac{N_2}{D_2}=\cdots=\frac{N_i}{D_i}=\cdots=\frac{\sum_iN_i}{\sum_iD_i} \nonumber
\end{eqnarray} we finally get
\begin{eqnarray}
\alpha(\epsilon,\mu_{\uparrow}-\mu_{\downarrow}) &=& \exp\left\{-\frac{\epsilon+\mu_{\uparrow}-\mu_{\downarrow}}{k_BT}\right\} \nonumber
\end{eqnarray} From the above relation it trivially follows that
\begin{eqnarray}
\alpha(\epsilon,\mu_{\uparrow}-\mu_{\downarrow}) &=& \alpha(\epsilon+\mu_{\uparrow}-\mu_{\downarrow},0)
\label{eq650}
\end{eqnarray} From the above relation and the relation in eq.~\ref{eq450} (which shows that $\langle S^{Mn}_z\rangle$ at steady state is solely a function of $\alpha$) we arrive at the result in eq.~\ref{Mn_spin_eq}

\end{document}